\documentclass[fleqn,10pt]{wlscirep}
\title{Screened moments in a Kondo insulator}

\usepackage{graphicx}
\usepackage{lipsum}
\usepackage{capt-of}

\author[1,*]{W. T. Fuhrman}
\author[1,2]{J. R. Chamorro} 
\author[3,4]{P. Alekseev}
\author[5]{J.-M. Mignot}
\author[6,7]{T. Keller}
\author[1, 8]{P. Nikolić}
\author[1,2, 9]{T. M. McQueen}
\author[1, 9]{C. L. Broholm}
\affil[1]{Institute for Quantum Matter and Department of Physics and Astronomy, The Johns Hopkins University, Baltimore, Maryland 21218 USA}
\affil[2]{Department of Chemistry, The Johns Hopkins University, Baltimore, Maryland 21218 USA}
\affil[3]{National Research Centre ``Kurchatov Institute", 123182 Moscow, Russia}
\affil[4]{National Research Nuclear University ``MEPhI", 115409 Moscow, Russia}
\affil[5]{Laboratoire L\'{e}on Brillouin, CEA-CNRS, CEA/Saclay, France}
\affil[6]{Max Planck-Institut für Festkörperforschung, Heisenbergstrasse 1, D-70569 Stuttgart, Germany}
\affil[7]{Max Planck Society Outstation at the Forschungsneutronenquelle Heinz Maier-Leibnitz (MLZ), D-85747 Garching, Germany}
\affil[8]{School of Physics, Astronomy and Computational Sciences, George Mason University, Fairfax, Virginia 22030, USA}
\affil[9]{Department of Materials Science and Engineering, The Johns Hopkins University, Baltimore, Maryland 21218 USA}
\affil[*]{wes@jhu.edu}

\begin{document}

\maketitle

\noindent \textbf{The Kondo impurity effect in metals is the result of quantum-mechanical scattering of conduction electrons by magnetic impurities.\cite{kondo1964resistance} This process leads to exotic effects such as unconventional superconductivity and an effective electron mass enhanced by 2-3 orders of magnitude.\cite{stewart1984heavy, mathur1998magnetically} Long known to have thermodynamic properties at odds with being an insulator, SmB$_6$ was recently found to exhibit bulk de Haas--van Alphen oscillations that were previously thought to be exclusively associated with metals.\cite{geballe1970properties, laurita2016anomalous, tan2015unconventional} We have observed a moment-screening effect in nominally pure and Gd-doped SmB$_6$ via heat capacity, magnetization, and resistivity measurements, and show this Kondo-like effect provides a natural explanation for metal-like phenomena in insulators stemming from the strongly interacting host system. This work extends the long-standing knowledge of the Kondo impurity effect to insulating systems and resolves decades of mysteries in the strongly-correlated topological insulator SmB$_6$.\cite{tan2015unconventional,laurita2016anomalous, wolgast2013low, kim2014topological,  fuhrman2015interaction} The requisite interactions are not unique to SmB$_6$, suggesting our discovery will provide an understanding of metallic properties in correlated insulators and new opportunities for probing their band structures.}

A mainstay of condensed matter physics, mixed valent SmB$_6$ is now understood to be a strongly correlated topological insulator.\cite{wolgast2013low} The low-energy regime has been examined extensively, exposing exotic phenomena such as bulk AC conduction, quantum oscillations, and magnetic quantum  criticality,\cite{laurita2016anomalous, tan2015unconventional,  akintola2017quantum} that have evoked even more exotic explanations.\cite{ PhysRevLett.115.146401, baskaran2015majorana} SmB$_6$ is itself described as a Kondo \emph{lattice}, where electrons from different orbitals hybridize non-locally to form a strongly-correlated insulator. The Kondo \emph{impurity} effect known in metals requires a finite density of states (DOS) at the Fermi energy (E$_F$) and localized magnetic moments. We propose that Gd doping simultaneously creates gapless degrees of freedom and localized magnetic impurities which together result in a dynamically screened Gd moment analogous to the Kondo impurity effect in metals.

Research into the role of impurities in SmB$_6$ spans decades.\cite{geballe1970properties, kasuya1979valence, gabani2002investigation, kim2014topological, PhysRevX.4.031012} Gd impurities in SmB$_6$ are electron donors similar to La, the non-magnetic lanthanide which acts as a Kondo hole, so 4$f^7$ Gd$^{3+}$ can be expected to act as a spinful Kondo hole.\cite{konovalova1982effect, schlottmann1992impurity} However, EPR studies of Gd doped SmB$_6$ show a strong negative g-shift and suggest that Gd impurities host an additional electron within a 4$f^7$ 5$d^1$ state for temperatures below 6 K.\cite{kunii1985electron, wiese1990possible}  La doping produces an approximately T-independent Sommerfield coefficient and increased susceptibility.\cite{nefedova1999imperfection,gabani2002magnetic} We observed $\gamma = $40 mJ/mol-K$^2$ at 5\% La, similar to previous studies of electron-doping via carbon impurities.\cite{PhysRevX.4.031012} This is consistent with theory which predicts Kondo holes produce an imaginary component in the $f$ electron self-energy resulting in a T-linear term in heat capacity and a Pauli-paramagnetic DOS at E$_F$.\cite{schlottmann1992impurity,nefedova1999imperfection,gabani2002magnetic} 

Quantifying the amount and effects of dilute impurities in SmB$_6$ is difficult. Rare earth elemental purification and impurity analysis is highly nuanced as there are both spectroscopic and isotopic mass overlaps. Nonetheless, semi-quantitative analysis is informative, and previous undoped samples show $>0.25\%$ rare earth and alkaline earth impurities.\cite{phelan2016chemistry} The low-temperature rise in linear specific heat in SmB$_6$ has been claimed intrinsic, appearing more pronounced but with lower magnitude in an isotopically pure sample.\cite{gabani2002investigation} However, $^{154}$Sm mass purification also retains $^{154}$Gd, the only other stable isotope with mass 154. Such isotopic samples present a unique look at the dilute limit of Gd impurities in otherwise exceptionally clean samples. 

\begin{figure}
\begin{center}
\includegraphics[totalheight=.8\textheight]{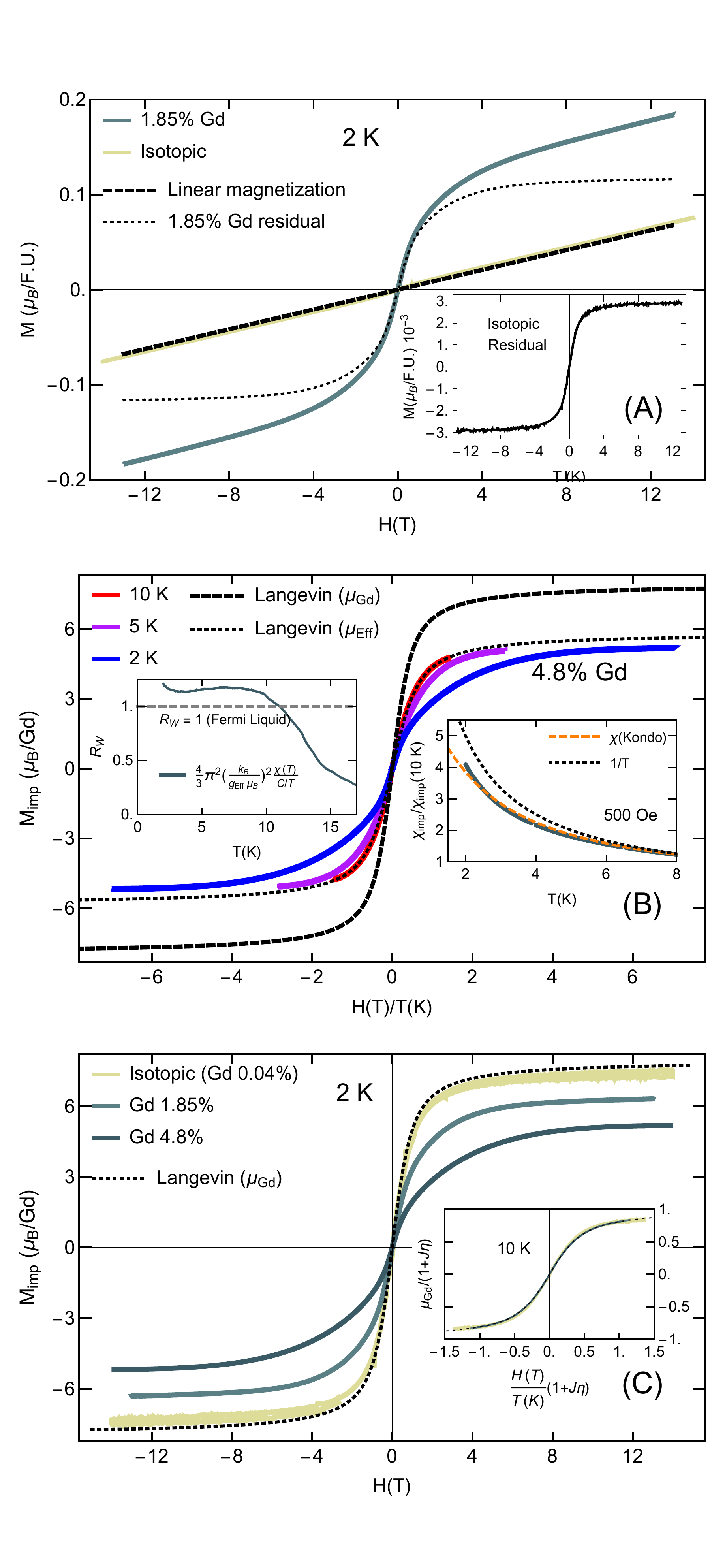}
	\caption{\label{Magnetization} Magnetization of SmB$_6$.  (A) Raw magnetization of 1.85\% Gd doped sample and isotopic $^{154}$Sm$^{11}$B$_6$. We ascribe the linear contribution(dashed) to pure SmB$_6$ and the residual magnetization (dotted line, 1.85\%) to impurities. Inset shows residual magnetization of the isotopic sample. (B) Residual magnetization of 4.8\% Gd doped samples at 2 K, 5 K, and 10 K, vs H/T. Insets show the WR becoming approximately constant at low temperatures and the similarity of $\chi(T)$ to the Kondo impurity model rather than a Curie-like susceptibility (SI). $g_{\rm Eff}$ in the WR was determined by $g_{\rm Eff}\sqrt[]{J(J+1)} = \mu_{\rm Eff}$. (C) Residual magnetization at 2 K of samples with varying doping levels. Inset shows that at 10 K a data collapse is achieved when scaling $M_{\rm Imp}$ and $H/T$ with (1+$J\eta$) and 1/(1+$J\eta$), respectively.
    }
\end{center}
\end{figure}

The magnetic field scale associated with SmB$_6$ is very large. Magnetoresistance measurements indicate gap closure beyond 80 T, and high-field magnetization shows predominantly van Vleck-like  linear dependence on field to at least 60 T.\cite{cooley1999high, tan2015unconventional} This contribution obscures impurity physics, but after its removal the residual magnetization is strikingly paramagnetic. We build our analysis on this component. Fig.1(A) shows the magnetization of the 1.85\% Gd and isotopic ($^{154}$Sm$^{11}$B$_6$) samples at 2 K. For Gd$^{3+}$, S = 7/2 ($\mu_{Gd} = 7.94\mu_B$) and the classical limit for magnetization is applicable. The Langevin function describing paramagnetic magnetization is determined by the effective magnetic moment $\mu_{\rm Eff}$ (field dependence), and concentration $c$ (overall amplitude). Fitting the residual magnetization curves at 10 K yields effective moments and concentrations of (7.74$\mu_B$, 0.04\%), ( 6.95$\mu_B$, 1.85\%), and (5.84$\mu_B$, 4.8\%) for  isotopic and nominally 2\% and 5\% Gd samples. Moment screening is unexpected for isolated Gd$^{3+}$ given the half-filled 4$f^7$ electron configuration, which carries no orbital moment and is particularly stable.\cite{zink2002magnetic} The applicability of the Langevin function for the rescaled residual magnetization of all samples at 10 K (Fig.1(C) inset) is evidence that a reduced-moment model accurately describes the high-temperature behavior for widely varying impurity concentrations. In addition to the reduced effective moment, we observe a departure from the paramagnetic Langevin function in the field-dependence of the magnetization which is most evident in the 4.8\% sample (Fig.1(B)). The temperature range for this deviation coincides with an uptick in the linear portion of the specific heat, which is greatly enhanced with doping (Fig.2(A)).

The observed magnetization is evidence of doped magnetic moments. In zero field, isolated moments cannot store heat, yet we observe a non-thermally-activated contribution to specific heat dependent on doping concentration. We conclude that Gd moments are coupled to additional degrees of freedom with a gapless spectrum. Although SmB$_6$ is non-metallic, we will show that the Kondo impurity model provides a qualitative description of the magnetization and specific heat data in SmB$_6$. Sm adjacent and exchange-coupled to Gd$^{3+}$ impurity sites would have a reduced valence. The intermediate valence fluctuations intrinsic to SmB$_6$ may then facilitate electronic fluctuations, screening localized Gd moments and mimicking the well-known Kondo impurity effect.  

\begin{figure}[!ht]
\begin{center}
\includegraphics[totalheight=.5\textheight]{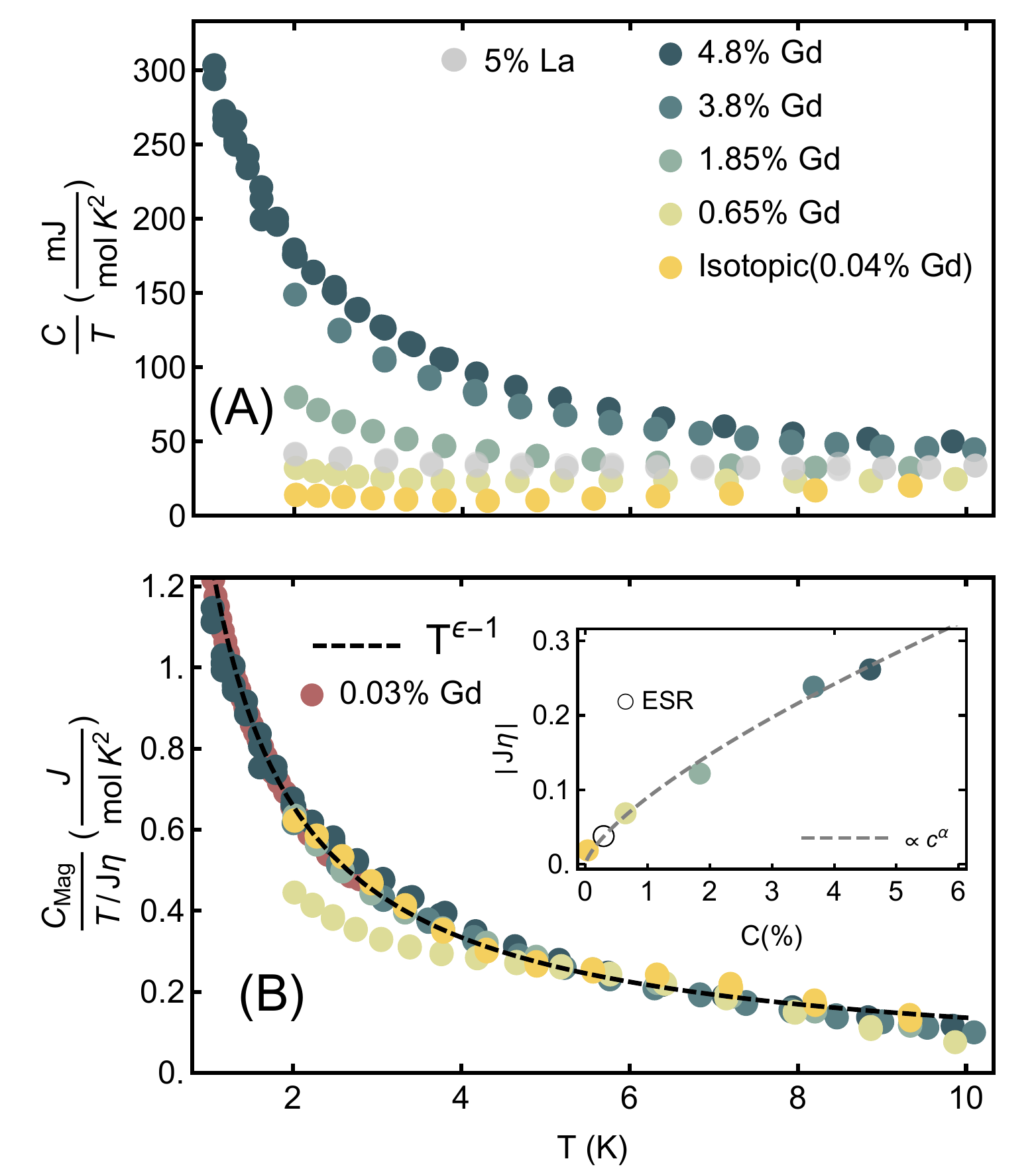} 
\caption{\label{HC} Heat Capacity of Gd and La-doped SmB$_6$.  (A) Raw heat capacity. The uptick in the linear portion of specific heat is most dramatic in the heavily Gd-doped samples, while 5\% La introduces a predominantly linear heat capacity at low temperatures at much lower magnitude per impurity than Gd doping. (B) Heat capacity of Gd-doped samples with lattice contribution ($\beta T^3$) removed, then scaled by $J\eta$ as determined from magnetization data. The same value of $\beta = 0.2$ (mJ K$^{-4}$ mol$^{-1}$) is used for all samples, obtained from ref.\cite{PhysRevX.4.031012}. The reduction in moment from g-factor shift seen via ESR measurements is included as an open circle.\cite{kunii1985electron} Previously published results for heat capacity of isotopic SmB$_6$  at low temperatures are scaled and included as red circles.\cite{gabani2002investigation}
}
\end{center}
\end{figure}

In the $s-d$ Kondo impurity model, magnetization and susceptibility are renormalized by $J\eta$, a dimensionless parameter formed by the product of an exchange constant $J$ and DOS $\eta$ (SI).  To first order in $J\eta$, the temperature independent correction to magnetization is $\mu_{\rm Eff} = \mu_{Gd}(1+J\eta)$ (see eq.(12) in SI). The effective moments extracted at 10 K can thus be used to estimate $J\eta$ (inset of Fig.2(B)). The Kondo susceptibility is given by $\chi = \chi_0(1+J\eta(1-J\eta log\frac{k_B T}{D})^{-1})$, where $D$ is a half-bandwidth of the band providing the DOS at Ef.\cite{yosida1996theory} Using $J\eta$ from the magnetization fits, the susceptibility of the 4.8\% Gd-doped sample gives $D = 0.7(2)$ meV. This modifies the 1/T susceptibility, which is also seen in the deviation from the Langevin function at low T (Fig 1(B)). In the Kondo model, this is a higher-order effect which indicates enhancement of Kondo screening at low temperatures. 

After subtracting the lattice contribution ($\beta T^3,  \beta = 0.2$ (mJ K$^{-4}$ mol$^{-1}$)), we scaled the specific heat by 1/$|J\eta|$, Fig.2(B). For a constant $J$, this amounts to scaling by the moment screening DOS. The data clearly converges to a single curve (the lightly doped but not isotopically purified 0.6\% sample may not be dominated by Gd impurities). Such scaling indicates the leading interactions are independent of doping, which is inconsistent with an origin in interactions between impurities.  The scaled specific heat has power-law dependence, $T^\epsilon$, where $\epsilon = .02(1)$, Fig.2(B). Previously published specific heat data can be scaled to follow the same power law fit to ~0.4 K, below which the power law fit diverges above the measured specific heat.\cite{gabani2002investigation} To first order in $\epsilon$, $T^\epsilon = 1-\epsilon log(T)$, so the data is qualitatively consistent with the Kondo impurity model $C_{Kondo} \propto 1+4J\eta log(k_B T/d)$, where $J\eta<0$ indicates a reduction in effective moment as observed.\cite{yosida1996theory}  However, $s-d$ Kondo models have corrections to specific heat with amplitude $\propto (J\eta)^4$ whereas Fig. 2B shows the specific heat scales as $C\propto J\eta$. Thus, our observed specific heat and it's relation to the magnetization is indicative of a new and distinct Kondo-like effect where low-energy degrees of freedom are themselves introduced by doping. The scaling collapse of specific heat ties the low temperature specific heat to the reduced impurity moment. Even the the low-temperature specific heat of our high purity $^{154}$Sm$^{11}$B$_6$ sample is fully accounted for by this impurity component. Fig.2(B) shows $J\eta$ scales approximately with $c^\alpha$, where $\alpha = 0.7(1)$. For comparison, theory predicts Kondo holes induce $\eta\propto \sqrt{c}$.\cite{schlottmann1992impurity}

For Fermi liquids, The Wilson ratio (WR) (inset of Fig.1(B)) of the spin susceptibility and and the specific heat over temperature tends to a constant in the low-T limit that is proportional to a Fermi liquid parameter.\cite{zou1986effective} For the impurity concentrations in SmB$_6$, we find it approaches a constant near unity, which adds another metal-like property. But since all our samples have exponentially activated DC resistivity, charge is ultimately localized despite the Fermi-liquid like WR. 

\begin{figure}[!ht]
\begin{center}
\includegraphics[totalheight=.5\textheight]{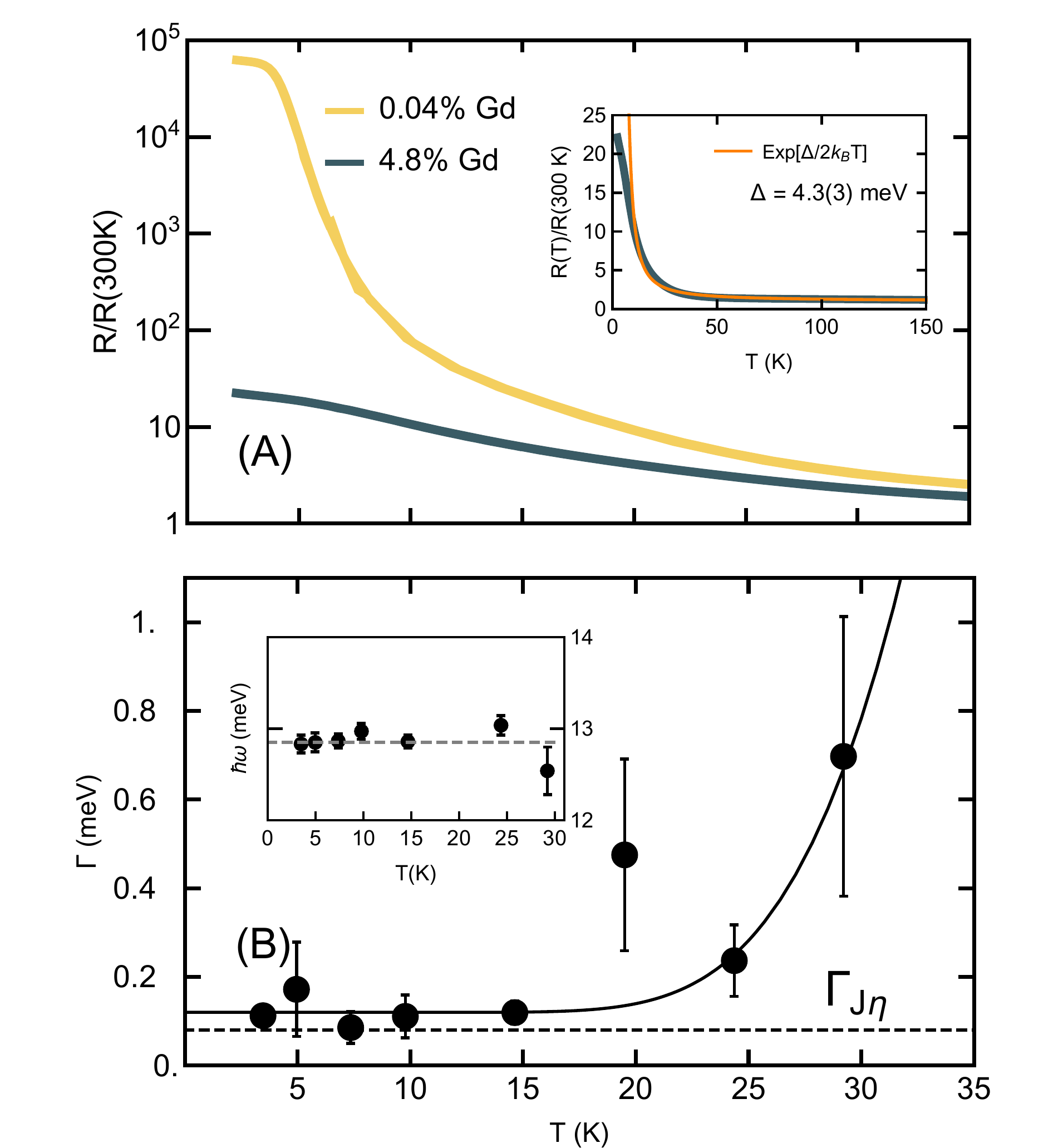} 
\caption{\label{RvT} Normalized resistance and exciton lifetime.  (A) Resistance of 4.8\% Gd doped (5.5 mm x 1 mm x1 mm rod) and isotopically purified $^{154}$Sm$^{11}$B$_6$ (2 mm x 2 mm x 1 mm chip). In the doped sample, R increases monotonically with decreasing T.  Inset shows R(T) for the 4.8\% Gd sample on a linear scale. Above 10 K R(T) is exponentially activated with a gap of 4.3(4) meV. (B) Spin-exciton lifetime and mode energy vs T. Solid line shows a non-thermal lifetime of $\Gamma_0 = 0.11 $meV and temperature dependence from exponential activation to the mode energy. $\Gamma_{J\eta}$ indicates the lifetime estimated from coupling the collective mode to the DOS at E$_F$ with the assumption of a common coupling constant for the exciton and Gd to the Kondo hole band.}
\end{center}
\end{figure}

The long but finite lifetime of the spin-exciton in SmB$_6$ (Fig.3(B)) is also connected to impurities. The spin-exciton arises from the coherent Kondo lattice, with a non-trivial theoretically infinite lifetime imbued by interactions.\cite{fuhrman2014gap} In an analogy to spin-resonance modes of cuprate superconductors,  coupling of the exciton to a Fermionic DOS at $E_F$ leads to a relaxation rate $\Gamma = 4\pi(g \eta )^2\Omega$, where $g$ is an exchange interaction ($g\propto t^2/U$ for the Hubbard model) and $\eta$ is the DOS.\cite{kee2002spin} Given the relatively flat 4$f$ bands in Sm$B_6$, coupling of the exciton to $\eta$ may be similar to the Kondo coupling $J$, and so $J\eta$ provides an estimate of $g \eta$. Substituting $g\eta = J\eta_{isotopic} = -.0227$ and the mode energy $\Omega = 12.8$ meV yields $\Gamma$ = 0.080 meV, which is similar to the observed value of 0.11 meV. The temperature dependence of the lifetime from 3.5 K to 15 K suggests only exponential activation, without evidence of coupling to additional energy scales. A dramatic effect on the spin-resonance mode and thermopower in the Tm-doped Kondo insulator YbB$_{12}$ has also been reported.\cite{alekseev2014possible} Such studies are also warranted for Gd-doped SmB$_6$. 

This moment-screening effect may manifest in other measurements. Valence Landau levels would periodically enhance the low energy DOS (LEDOS), giving rise to dHvA but not SdH oscillations via impurity moment screening with periodicity set by the Pseudo Fermi Surface (PFS) of the sharply dispersing $d$ bands.\cite{tan2015unconventional, PhysRevLett.115.146401,fuhrman2015interaction} Our susceptibility data shows that at low temperatures, higher order corrections enhance Kondo screening.  This would increase the screening of impurities by the LEDOS, leading to an increased amplitude at low temperatures and deviation from Lifshitz Kosevich behavior. Quantum oscillations of Gd doped Kondo insulators could reveal the PFS, which itself is defined by band inversion in Kondo insulators, the locations of which fully specify topological invariants in inversion-symmetric Kondo insulators.\cite{fu2007topological} The substantial DOS indicated by specific heat and the approximately constant WR is also consistent with the  optical conductivity which, although large, is orders of magnitude smaller than for heavy Fermion metals.\cite{laurita2016anomalous}

The profound sample variation and metal-like properties of SmB$_6$ find a coherent explanation in terms of a Kondo-like impurity moment-screening effect. An array of consequences of even dilute impurities, which are extremely difficult to avoid, are evident. The scaling of the low-temperature heat capacity resolves a mystery 40+ years in the making and the effect we introduce offers new routes to understanding and experimentally probing not only SmB$_6$, but strongly correlated insulators in general. It also provides a simple prescription for determining topological invariants via impurity dHvA oscillations and guides further studies into the intrinsic physics within perennially intriguing SmB$_6$.

\section*{Methods}
\subsection*{Crystal Growth}
Targeted 1\%, 2\%, and 5\% samples were grown by the aluminum flux growth method in which elemental Sm (Ames Laboratory), Gd (Ames Laboratory), La (Ames Laboratory), B (Alfa Aesar 99.99\%), and Al (NOAH Technologies 99.99\%) were added to an alumina crucible in a 0.005$-x$:$x$:0.03:3 molar ratio, with $x$ set to the desired (Gd, La) content. Al was premelted in the crucible prior to combining target elements and adding  additional Al, which was then heated to 1350$^\circ$C at a ramp rate of 100$^\circ$/hr. The temperature was held for 12 hours and then allowed to cool to 800$^\circ$C at a rate of 4.5$^\circ$C/hr, followed by cooling to room temperature at 100$^\circ$/hr. The aluminum flux was etched off with a caustic NaOH solution, and the crystals were isolated via vacuum filtration. Resultant crystals were rod shaped with typical dimensions up to 11 mm x 1 mm x 1 mm. The isotopic $^{154}$Sm$^{11}$B$_6$ sample was grown by Yu Paderno at IPMS (Kiev) by the  floating zone method and is the same sample used in other neutron spectroscopy studies.\cite{fuhrman2015interaction}

\subsection*{Measurements}
In-house powder x-ray diffraction measurements were performed to determine phase purity using Cu $\textit{K}\alpha$ radiation on a Bruker D8 Focus diffractometer with a LynxEye detector. Physical properties characterization was performed on a Quantum Design Physical Properties Measurement System (PPMS). Inelastic neutron scattering was performed on the TRISP hybrid triple-axis/neutron spin-echo spectrometer. We utilized a PG(002) monochromator and velocity selector with spin-echo coils in scanning DC operation for an estimated resolution of 10 $\mu eV$, which was subtracted in quadrature from the measured width to obtain the physical width $\Gamma$. Details of the instrument and technique are complex and are summarized elsewhere.\cite{keller2002nrse} 

\section*{Acknowledgements}
The work at IQM was supported by the US Department of Energy, office of Basic Energy Sciences, Division of Material Sciences and Engineering under grant DE-FG02-08ER46544. W.T.F. is grateful to the ARCS foundation, Lockheed Martin, and KPMG for the partial support of this work.  The authors thank Piers Coleman, Natalia Drichko, Seyed Koohpayeh, Nicholas Laurita, and W. Adam Phelan for engaging discussions. 

\section*{Author contributions statement}

W.T.F. conceived the idea of the experiment. W.T.F. and J.R.C. grew the crystals, measured properties, and performed data analysis.  T.K., W.T.F., and C.L.B. collected and analyzed the neutron scattering data. P.A. and J.-M.M. provided the doubly isotopically enriched $^{154}$Sm$^{11}$B$_{6}$ crystal. P.N. calculated thermodynamic quantities in the s-d Kondo model. All authors contributed to the manuscript.

\section*{Additional information}

The authors declare no competing financial interests. Supplementary information is available online. Correspondence and requests for materials should be addressed to W.T.F.

 \bibliography{SmB6_HC2.bib}
\end{document}